# Design and Fabrication of an Ultra-low Noise Ag-AgCl Electrode

B. Rostami, S.I. Mirzaei, A. Zamani, A. Simchi, M. Fardmanesh

*Abstract*— **We have developed a new method for fabrication of ultra-low noise silver-silver chloride (Ag-AgCl) reference electrode with dramatically high stability, useful in biomedical precise measurements. Low noise level and high stability are the most important features when evaluating reference electrode for example when used as a half-cell at electrochemical cells. The noise was measured for bare Ag-AgCl electrode pairs in a saline electrolyte without any junction. The noise spectral density for such measurements increases at low frequencies due to flicker (1/f) noise. Thermal noise from the real part of the electrode impedance is always lower than 1/f noise and is shown as offset voltage in noise spectrum. Electrode noise is highly correlated to its offset voltage. Moreover, the variance of the potential difference in our measurements is correlated to the noise level of electrodes. In this paper, offset voltage spectrum and the variance of data while doing measurements are proposed for identification of noise in our electrodes. The results presented suggest a new process for fabrication of electrode pairs with differential noise potential of 16 nanovolt root mean square at frequency of 1 Hz.**

## I. INTRODUCTION

In an electrochemical system, reference electrodes are essential components since they can provide stable and constant potential during the measurement [1], [2]. A major effort toward the development of reference electrodes for in vivo biosensing has been directed during the past half a century [3]. Many electrochemists frequently utilize the silver-silver chloride electrodes (Ag-AgCl) as the reference electrodes in their systems. This type of electrode is commonly used because of environmental reasons and its advantageous characteristics such as low cost, high stability and reproducibility of potential, and electrochemical reversibility [4], [5]. This type of electrode has attracted more and more attention for recording physiological electrical signals, including electroencephalography (EEG), electromyography (EMG), and electrocardiography (ECG) recording [6].

One method to measure the noise power spectrum density of the electrode pairs is to place them in saline electrolyte and measure the potential difference between them in an electromagnetically shielded environment [7-10]. The power spectral density of the output potential difference contains valuable data to determine the noise attributed to the electrodes. The power spectral density of differential potential measurements increases at low frequencies, which represents the flicker (1/f) noise in the system. Real part of the impedance of the electrodes contributes to the thermal noise. The offset voltage in the high frequency part of the power spectrum is limited by the thermal noise which is always lower than 1/f noise [11]. Electrode noise is highly correlated to its dc offset voltage [12]. When electrode contact area in the electrolyte increases, the impedance and therefore its thermal noise decreases [12].

It is shown that the microstructure of the Ag-AgCl has a large influence on its repeatability and performance [13]. One of the most effective microstructural features in electrodes is their porosity. More porosity in electrodes provides larger electrolytic contact area, enabling higher exchange current densities at equilibrium which reduces their polarization voltage, thereby resulting in a highly reproducible and stable reference potential [14], [15]. Thus, porous electrodes perform more effectively in numerous applications like sensors, batteries, and fuel cells [16], [17].

Some recent studies have focused on the fabrication of thermal electrolytic (ThE) Ag-AgCl reference electrodes [18-20]. ThE electrodes are conventionally prepared by applying $Ag_2O$ paste on a supporting wire skeleton (I. $Ag_2O$ paste deposition). This thin wire skeleton is usually made of a noble metal such as Platinum (Pt). Then $Ag_2O$ paste is thermally reduced to a porous Ag layer in furnace (II. Thermal annealing). Finally, 10-25 % of Ag layer is converted to AgCl through electrolysis (III. Anodization). The degree of porosity in ThE reference electrodes needs a critical optimization. High degree of porosity in ThE electrodes increases the probability of contact between the electrolyte and the wire skeleton. This occurrence will create mixed potentials as solution penetrates deeper [21-26]. Therefore, for increasing the exchange current densities at equilibrium which results in a highly reproducible reference potential, there is an upper limit for increasing surface porosity.

In this research we developed a new technique to increase the effective surface of ThE reference electrodes by using a light Rhodium (Rh) coated silver tube as the inner skeleton instead of Pt wire for supporting the Ag-AgCl layers. Rh is also a noble metal which has a similar electrochemical reaction mechanism as Pt wire. Since we deposit a very thin layer of Rh on the outer surface of Ag tube, this method is more cost effective than using Pt wire as used in the conventional ThE methods.

In the following sections, two different methods for fabrication of Ag-AgCl reference electrodes are proposed and their microstructure and noise level are compared to each other. The first method is the simplest way to fabricate an Ag-AgCl reference electrode by only doing the anodization step on a silver rod (Basic standard method) and the second one is the ThE method by replacing Pt wire by a light Rhodium (Rh) coated silver tube (New ThE method).

## II. BASIC STANDARD METHOD

In the basic standard fabrication method, a proportion of Ag rod was converted to AgCl by electrolysis technique which could be driven by applying a fixed current or a fixed potential. In this research, we have used constant current for converting 15 percent of the mass of identical Ag rods (99.99%, rectangular cross section 1mm x 0.5mm) to AgCl in a chloride solution. The block diagram of basic standard fabrication process is shown in Fig. 1.

Prior to electrode preparation, to remove surface contaminants of the silver rods, they were immersed in $HNO_3$ solution (0.1 M, 37%, for 5 minutes) and rinsed thoroughly with DI water. Afterwards, they were immersed in $NH_4OH$ solution (37%, for 5 minutes) to remove oxide surface contaminants and again rinsed thoroughly with DI water. Aqueous solutions of KCl (0.5 M) and HCl (0.004 M) were mixed and used as the electrolyte in electrolysis process. Prior to anodization, the mass of the part of each silver rod, which will be immersed in the electrolyte, is determined to calculate the amount of charge needed to convert 15 percent of that part of mass to AgCl.

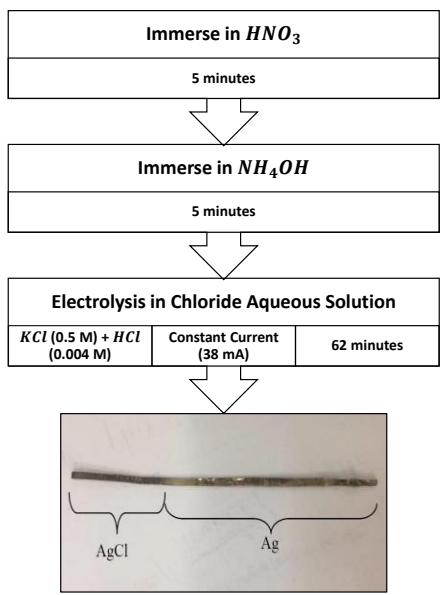

Fig. 1. Block diagram of the basic standard method for fabrication of Ag-AgCl electrodes.

## III. NEW ThE METHOD

In this method, Ag-AgCl electrodes were prepared by N times applying $Ag_2O$ paste to an Ag tube with Rh coating (99.99% Ag, 1 mm inner diameter and 1.6 mm outer diameter), followed by its thermal decomposition (100 °C for 30 minutes followed by 500 °C for 2 hours) each time. The block diagram of preparation steps of silver oxide paste is shown in Fig. 2.

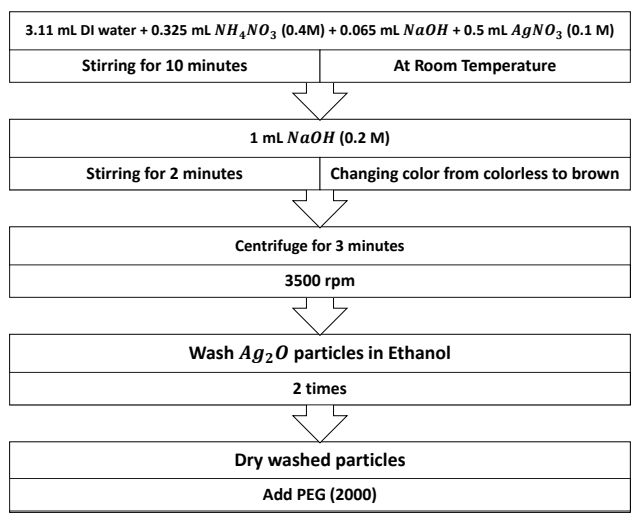

Fig. 2. Silver oxide paste preparation steps.

Synthesized Ag₂O paste was deposited on the Ag tubes with Rh coating and was thermally decomposed. This step was repeated N times to obtain a thick enough layer of porous Ag on the tube-shaped skeleton. Approximately 15 percent of the mass of the part of the porous Ag, which would be immersed in the electrolyte, was electrolytically converted to AgCl through anodization with constant potential in an aqueous solution of HCl (0.1 M). Finally, the prepared electrodes were stored in a KCl solution (3 M). The general block diagram of the new ThE method for fabricating Ag-AgCl reference electrode is shown in Fig. 3. Two different pairs of this type of electrodes with different N (One of them with N=1 and the other one with N=3) were processed.

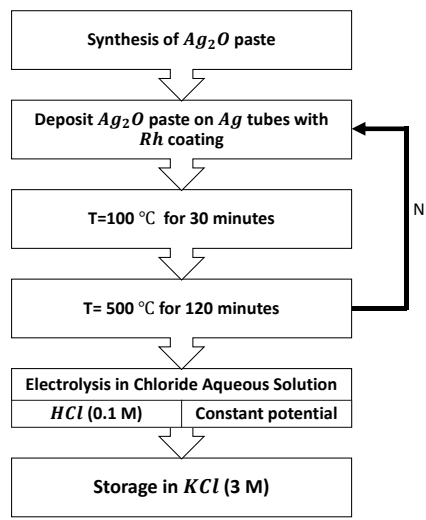

Fig. 3. Block diagram of the silver paste method for fabrication of Ag/AgCl electrodes.

## IV. EXPERIMENTAL RESULTS AND DISCUSSION

In this section, the noise of four different pairs of Ag-AgCl electrodes are compared to each other. The first pair were fabricated by using the basic standard method. Second and third pairs were processed by using the new ThE method with N=3 and N=1, respectively. The fourth pair are commercial double-junction Ag-AgCl reference electrodes. First, the microstructure of two proposed methods are compared by studying their Scanning electron microscopy (SEM) images. Next, the differential voltage noise of the four electrode pairs are compared.

SEM measurements were carried out to study the effect of different fabrication methods on the microstructure of Ag-AgCl reference electrodes. The SEM images of the surface of the electrodes fabricated by the new proposed ThE method and the basic standards method are shown in Fig. 4. As it is shown, both proposed methods result in porous microstructures, however the electrode prepared with the new ThE method (Fig. 4 (b)), appears to be mechanically very rigid with larger pores. A partial explanation for these structural differences could be the fact that in the silver oxide paste method the annealing step for 1 hour at 100∘C, allows for the creation of larger holes in microstructure which is due to the removal of PEG molecules.

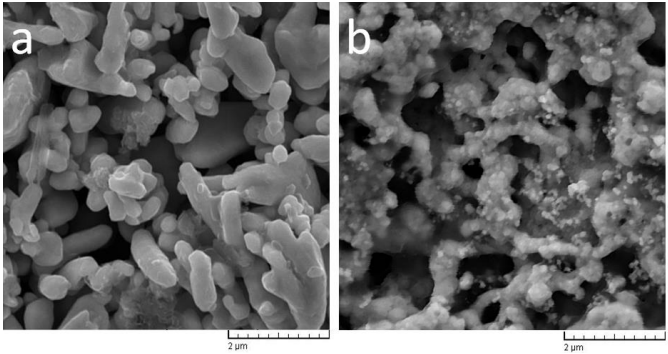

Fig. 4. Scanning electron microscopy measurement: (a) basic standard method, (b) new ThE method.

In this research the noise quantification was achieved by measuring differential output potentials of two identical electrodes in a chloride solution in an electromagnetically shielded box. The electrolyte used for all the differential noise level experiments was aqueous solution of NaCl (3 M). In this measurement, both intrinsic noise (due to the different microstructures and not identical electrodes) and system noise (due to the noise of output readout and amplification circuits) were measured together.

During the noise test, temperature, humidity and air pressure inside the electromagnetically shielded box were measured. Since all the measurements were done in almost same environmental conditions (such as constant temperature, humidity, and air pressure), system noise was assumed to be the same for all measurements. The noise spectrum of system, captured for 2000 seconds, is shown in Fig. 5. As it is shown, the system noise level over a wide frequency range is almost constant and around 0.45 nV/√Hz.

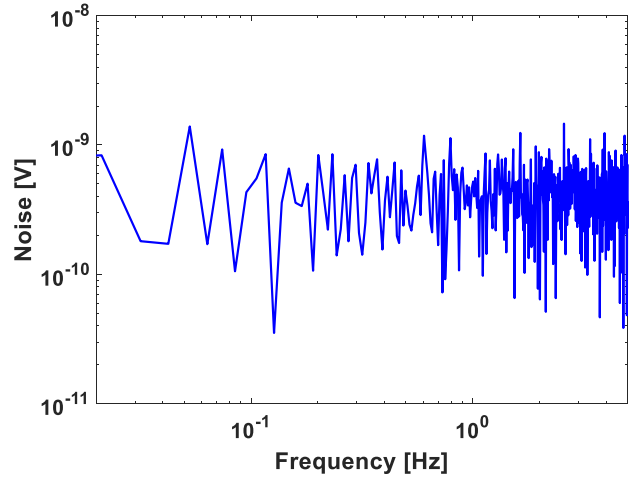

Fig. 5. Noise spectrum of the system over 2000 seconds.

The noise level of electrodes immediately after inserting them in the electrolyte is much different from when keeping them in the same electrolyte for a while. This is due to imbalance between the solution ions and surface molecules of electrodes which affects the half-cell potential of electrodes. Therefore, it takes a while to reach an equilibrium condition on the surface of electrodes in electrolyte. It takes at least 48 hours to reach a relatively acceptable balance in the measured noise level. Fig. 6 shows how the differential output potential stabilizes during the measurement time.

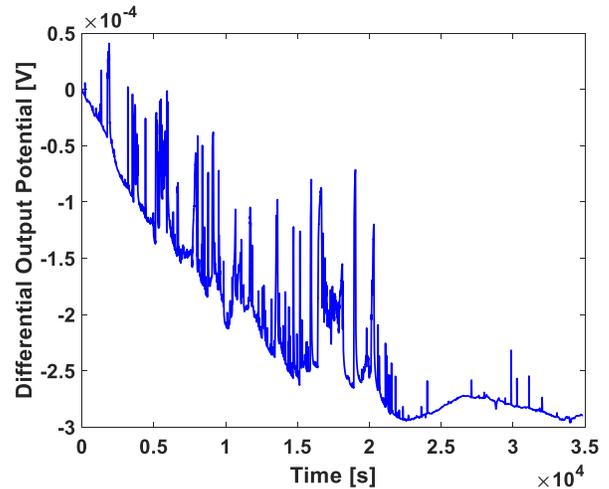

Fig. 6. Differential output measurement stabilization during the measurement time.

In this part, the differential noise level of four different pairs of Ag-AgCl electrodes are compared. The differential noise level of each pair was monitored over many days after fabrication, keeping all electrodes under same conditions, until they reached a relatively stable noise level. The first pair were fabricated by using the basic standard method (named A) and their corresponding differential noise level was tested immediately after insertion to the electrolyte and after 10 days of storage in the same electrolyte. The noise spectrum of the recorded voltages, captured for 7000 seconds, is shown in Fig. 7. The noise level at 1 Hz decreases evenly from about 670 nV/√Hz to 84 nV/√Hz.

As mentioned in new ThE method section, to investigate the effect of the number of applied silver oxide layers (N in Fig. 3) and its post processing steps, two different pairs of Ag-AgCl electrodes (one pair with N=1 and the other pair with N=3) were fabricated. The second pair of electrodes with one layer of silver oxide (named B1), were tested immediately after insertion to the electrolyte, after one day, four days, eight days, and twelve days while kept in the same electrolyte. The noise spectrum of the recorded voltages, captured for 20,000 seconds, is shown in Fig. 8. The noise level at 1 Hz decreases evenly from 11 µV/√Hz to 16 nV/√Hz.

The third pair of electrodes with N=3 (named B3) were tested immediately after insertion to the electrolyte, after three days, six days, and ten days being kept in the same electrolyte. The noise spectrum of the recorded voltages, captured for 20,000 seconds, is shown in Fig. 8. The noise level at 1 Hz decreases generally from 5.5 µV/√ Hz at B3-immediately to 212 nV/√ Hz after 10 days. However, the decrease of noise level for this type of electrode is not uniform and increases after three days.

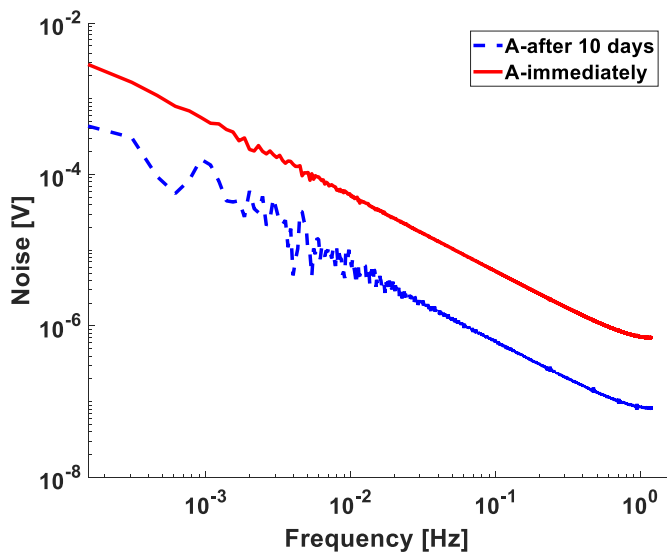

Fig. 7. Noise spectrum of the face to face differential voltage level of A type electrodes fabricated through the basic standard method over 7000 seconds.

A partial explanation for this observation could be the time degradation of the adhesion between silver oxide layers of the electrodes of type B3 in the electrolyte, which increases the effective impedance of the electrodes in the electrolyte, leading to an increase in its corresponding noise level.

One of the reasons for temporal degradation of the adhesion between silver oxide layers could be the fact that the three layers of deposited silver are not exactly similar to each other: the first silver oxide layer is heated in the furnace three times while the second layer is exposed to heat twice and the third layer is exposed to heat only once. Therefore, the disparity between different layers would be one of the most probable reasons for adhesion degradation between layers.

Another perception from Fig. 8 is that the noise level at 1 Hz, immediately after insertion, for type B3 is less than the corresponding noise level for B1, which is due to thicker layer of porous silver chloride in B3 compared to B1. This observation is verified by Electrochemical Impedance Spectroscopy (EIS) measurement comparing the impedance of electrodes type B1 and B3, conducted immediately after inserting them in electrolyte, which their corresponding results are shown as Bode plots and Nyquist diagrams in Fig. 9. As shown in Fig. 9, the real part of impedance of electrode type B3 is lower than that of electrode type B1, which is in harmony with the DC offset in noise spectrum density (Fig. 8, B1 immediately and B3 immediately).

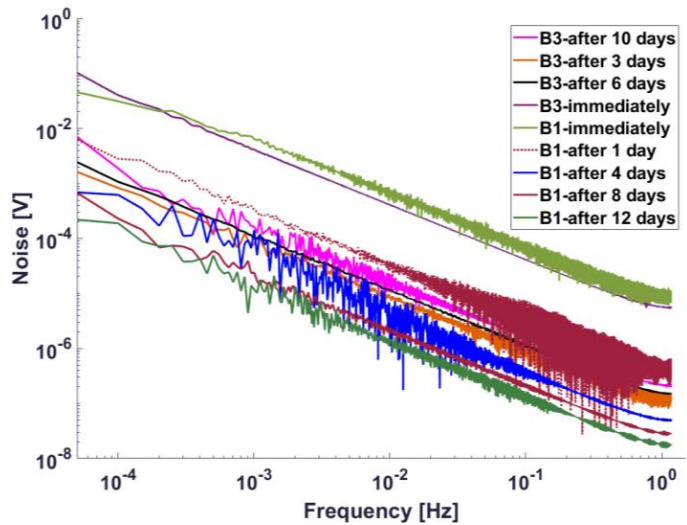

Fig. 8. Noise spectrum of the face to face differential voltage level of B1 and B3 type electrodes fabricated through the new ThE method over 20000 seconds.

The fourth pair for noise level test consists of commercial double-junction Ag-AgCl reference electrodes (named C). Their corresponding differential noise level was tested after 10 days of storage in the same electrolyte. The noise spectrum of the recorded voltages, captured for 7000 seconds, is shown in Fig. 10. The noise level at 1 Hz has reached a stable level of 1 µV/√ Hz after 10 days.

Other than the offset voltage in the noise spectrum, the variance of the recorded potential difference in face to face measurements is also correlated to the noise level of the electrodes. We calculated the variance of the recorded data while doing differential noise measurements and compared them for different pairs of electrodes. Fig. 11 shows a comparison between the variance of the electrode pairs type B1 and B3 which are fabricated using the new ThE method. As it is shown, the variance of the recorded data from electrodes type B1 after reaching its stable state (after 12 days) is around 3 pV. However, for B3 electrodes, this value goes down to 0.1 nV after 3 days and increases to 1.3 nV after reaching its stable state (after 10 days).

Fig. 12 shows a comparison between the variance of the electrode pairs type A which were fabricated using the basic standard method and C that are commercial double-junction electrodes. As it is shown, the variance of the recorded data from electrodes type A and C after reaching their steady state (after 10 days) are around 30 pV and 8 nV respectively.

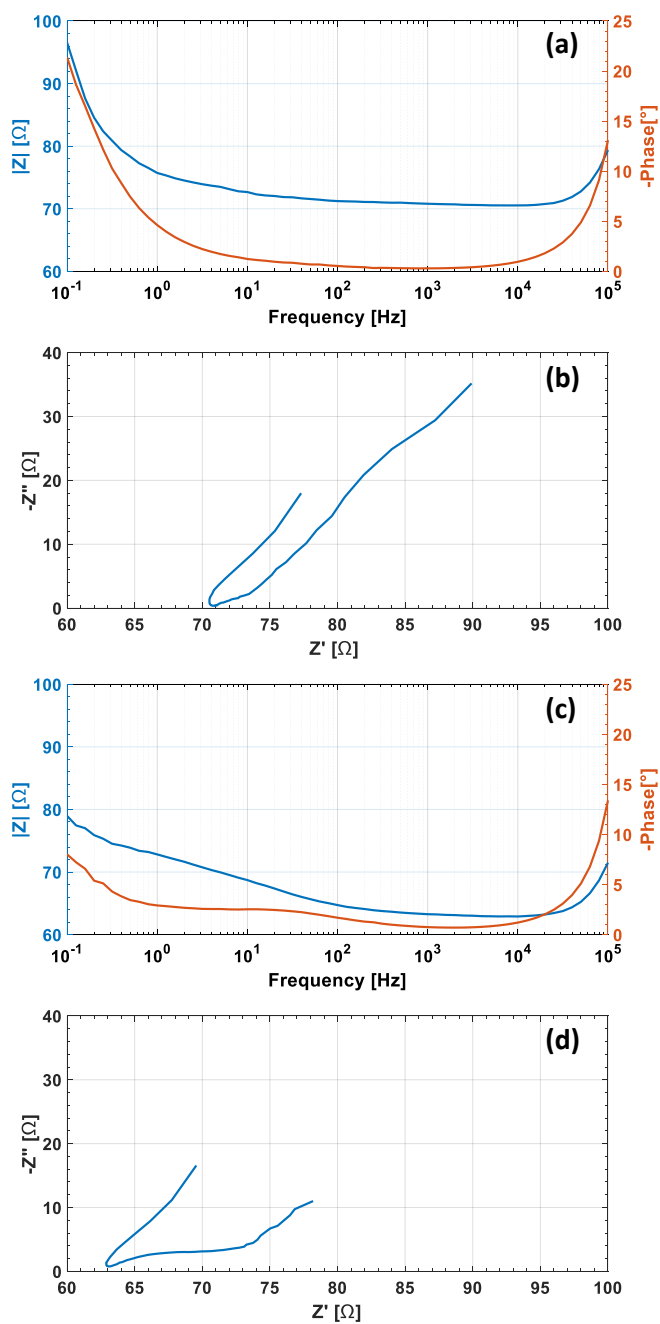

Fig. 9. EIS measurements immediately after inserting in the electrolyte; electrode type B1: (a) Bode plot (b) Nyquist diagram, and electrode type B3: (c) Bode plot (d) Nyquist diagram.

comparison with the commercial double-junction Ag-AgCl pair. The electrode pair type B1 has the lowest noise level among other types, which regards to its dc offset voltage in noise spectrum. At frequency of 1 Hz, its noise level at steady state is around 16 nV/√Hz which is acceptable among reported noise levels of reference Ag-AgCl electrodes so far.

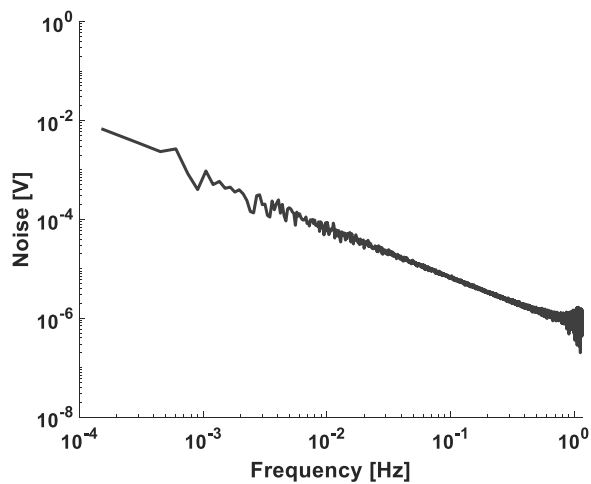

Fig. 10. Noise spectrum of the face to face differential voltage level of C type electrodes (the commercial double-junction Ag-AgCl reference electrode) over 7000 seconds.

TABLE I
THE MEASURED NOISE LEVELS OF FOUR ELECTRODE PAIRS AT THEIR STEADY STATES

| Noise Level at Steady State | A | B1 | B3 | C |
|---|---|---|---|---|
| DC offset voltage (nV @ 1 Hz) | 84 | 16 | 212 | 1000 |
| Data variance (nV) | 0.03 | 0.003 | 1.3 | 80 |

The measured noise levels of four electrode pairs at their steady states, resulted from the dc offset voltage in the noise spectrum and the variance of the recorded data while doing face to face noise measurements, are summarized in Table I. As it is shown, type B1 and type B3 have more porous and rigid structures in comparison with type A and this is the main reason for lower noise level in type B1 compared with type A.

Despite the more porous structure of type B3 against type A, the degradation of adhesion between three silver oxide layers in type B3 increases its noise level after a while being kept in the electrolyte solution. All fabricated electrodes through our proposed methods have much lower noise level in

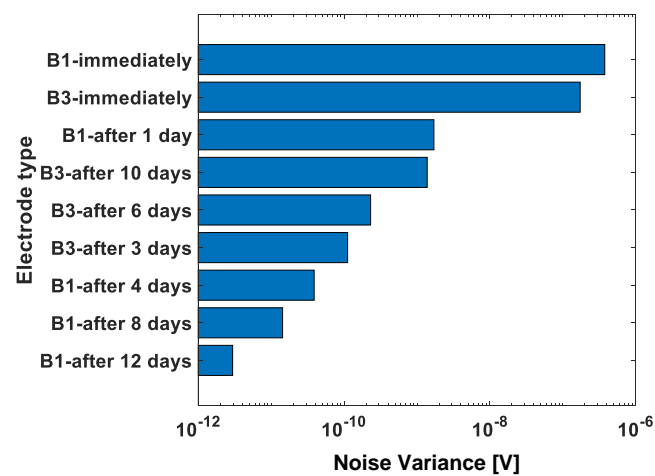

Fig. 11. A comparison between the variance of the electrode pairs type B1 and B3 which were fabricated through the new ThE method.

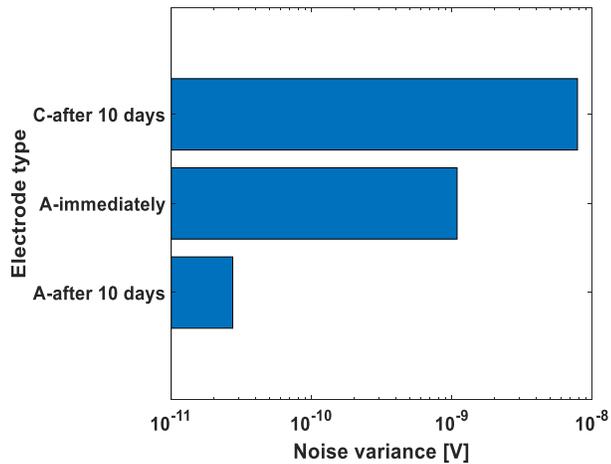

Fig. 12. A comparison between the variance of the electrode pairs type A which were fabricated through the basic standard method and C that are commercial double-junction Ag-AgCl electrodes

## V. CONCLUSION

In this paper we proposed two different techniques for fabrication of low noise silver-silver chloride (Ag-AgCl) reference electrodes that are widely utilized in biomedical precise measurements. To find the optimum approach to fabrication of an ultra-low noise reference electrode, the electrochemical and noise properties of four different pairs of electrodes were measured and compared to each other. The noise was characterized by measuring the differential potential from Ag-AgCl electrode pairs in aqueous NaCl (3 M) electrolyte. Electrode noise performance is highly correlated to its DC offset voltage in the noise spectrum and the variance of the recorded data, while performing differential measurements. In this paper, we demonstrated a new scheme for fabricating pairs of electrodes using the proposed novel ThE method with one layer silver oxide paste, resulting in 16 $nV_{RMS}$ noise voltage at 1 Hz and 3 $pV_{RMS}$ noise variance, measured while located face-to-face in electrolyte environment.